\documentclass[10pt,conference]{IEEEtran}
\IEEEoverridecommandlockouts
\usepackage{cite}
\usepackage{amsmath,amssymb,amsfonts}
\usepackage{algorithmic}
\usepackage{graphicx}
\usepackage{textcomp}
\usepackage{xcolor}
\usepackage{url}
\usepackage{booktabs}
\usepackage[most]{tcolorbox}
\usepackage[caption=false,font=footnotesize]{subfig}
\usepackage[T1]{fontenc}
\def\BibTeX{{\rm B\kern-.05em{\sc i\kern-.025em b}\kern-.08em
    T\kern-.1667em\lower.7ex\hbox{E}\kern-.125emX}}
\begin{document}

\title{Retrieval-Augmented Code Review Comment Generation\\
}

\author{
\IEEEauthorblockN{Hyunsun Hong}
\IEEEauthorblockA{\textit{School of Computing} \\
\textit{KAIST}\\
Daejeon, Republic of Korea \\
hyunsun.hong@kaist.ac.kr}
\and
\IEEEauthorblockN{Jongmoon Baik}
\IEEEauthorblockA{\textit{School of Computing} \\
\textit{KAIST}\\
Daejeon, Republic of Korea \\
jbaik@kaist.ac.kr}
}

\maketitle

\begin{abstract}
Automated Code review comment generation (RCG) aims to assist developers by automatically producing natural language feedback for code changes. Existing approaches are primarily either generation-based, using pretrained language models, or information retrieval-based (IR), reusing comments from similar past examples. While generation-based methods leverage code-specific pretraining on large code–natural language corpora to learn semantic relationships between code and natural language, they often struggle to generate low-frequency but semantically important tokens due to their probabilistic nature. In contrast, IR-based methods excel at recovering such rare tokens by copying from existing examples but lack flexibility in adapting to new code contexts—for example, when input code contains identifiers or structures not found in the retrieval database. To bridge the gap between generation-based and IR-based methods, this work proposes to leverage retrieval-augmented generation (RAG) for RCG by conditioning pretrained language models on retrieved code–review exemplars. By providing relevant examples that illustrate how similar code has been previously reviewed, the model is better guided to generate accurate review comments. Our evaluation on the Tufano et al. benchmark shows that RAG-based RCG outperforms both generation-based and IR-based RCG. It achieves up to +1.67\% higher exact match and +4.25\% higher BLEU scores compared to generation-based RCG. It also improves the generation of low-frequency ground-truth tokens by up to 24.01\%. We additionally find that performance improves as the number of retrieved exemplars increases.
\end{abstract}

\begin{IEEEkeywords}
Code Review Automation, Deep Learning, Software Quality Assurance, Retrieval-Augmented Generation
\end{IEEEkeywords}

\section{Introduction}

Code review is a systematic and collaborative process that involves examining code written by others to assess its quality and readiness for integration. If a code change is deemed unsatisfactory, the reviewer may reject the merge request and provide feedback in the form of review comments, enabling the author to revise the code accordingly. This process plays a key role in identifying latent issues early in the development cycle, preventing potential defects, and improving maintainability through refactoring \cite{mcintosh2016empirical}. For this reason, code review is widely adopted and strongly recommended practice as a core component of software quality assurance \cite{fregnan2022happens}. However, code review is inherently labor-intensive and time-consuming, requiring experienced developers to thoroughly inspect changes and deliver precise, constructive feedback \cite{huang2022reviewing}. Studies report that developers spend an average of six hours per week on review-related activities, including reading, understanding, and providing feedback on code changes \cite{bosu2013impact}. Moreover, authors often face delays of 15 to 64 hours before receiving review comments, which can hinder development velocity \cite{hong2022should}. As software systems grow, the volume of code review requests increases significantly; for example, Microsoft Bing reportedly conducts approximately 3,000 code reviews per month \cite{rigby2013convergent}. These challenges have led researchers to explore automating the review process. Review Comment Generation (RCG) has emerged as a promising approach, aiming to ease the burden of manual review comment writing by automatically generating relevant, context-aware feedback for code changes.

RCG has been researched in two directions: Generation-based RCG and Information-Retrieval (IR)-based RCG. Generation-based RCG utilizes a generative language model. With the hypothesis that software exhibits repetitive and predictable property \cite{hindle2016naturalness}, generation-based RCG takes advantage of probabilistic language models trained on the large code-review corpus to generate most probable code review words. Tufano et al. \cite{tufano2021towards, tufano2022using} were first to apply pre-trained language model (PLM) to code review automation and Li et al. \cite{li2022codereviewer} later proposed a code review-specific pre-training methodology. Both studies contributed benchmark datasets that have since become foundational for evaluating RCG methodology \cite{tufano2022using, li2022codereviewer}. Based on these benchmarks, subsequent works have focused on adapting advanced natural language processing (NLP) techniques to enhance generation-based RCG \cite{zhou2023generation, lu2023llama, nashaat2024towards}. IR-based RCG, on the other hand, utilizes information retrieval. It finds the most probable code reviews from a large code review history database and regards these retrieved exemplars as code reviews for the input code \cite{hong2022commentfinder, shuvo2023recommending, kartal2024automating}. Rather than generating new text from generative language model, IR-based RCG reuses exemplar mined from a code review history database. It is more efficient than generation-based RCG because it does not require any training \cite{hong2022commentfinder}. Notably, CommentFinder \cite{hong2022commentfinder}, an IR-based RCG approach, reports a 32\% performance improvement over the generation-based method proposed by Tufano et al. \cite{tufano2022using} in RCG. This enhancement is achieved with a 49× speedup over the generation-based method \cite{tufano2022using}, by leveraging simple Bag-of-Words lexical similarity to retrieve similar code examples from the training dataset. Although both generation-based RCG and IR-based RCG have been studied independently with its own advantages, their integration has not been explored. However, there is room for two methods to be integrated harmoniously.

As generative language models innately tend to generate high-frequency words, it has trouble with generating low-frequency but semantically important tokens—referred to as 
\begin{figure}[t]
\centering
\includegraphics[width=0.9\linewidth]{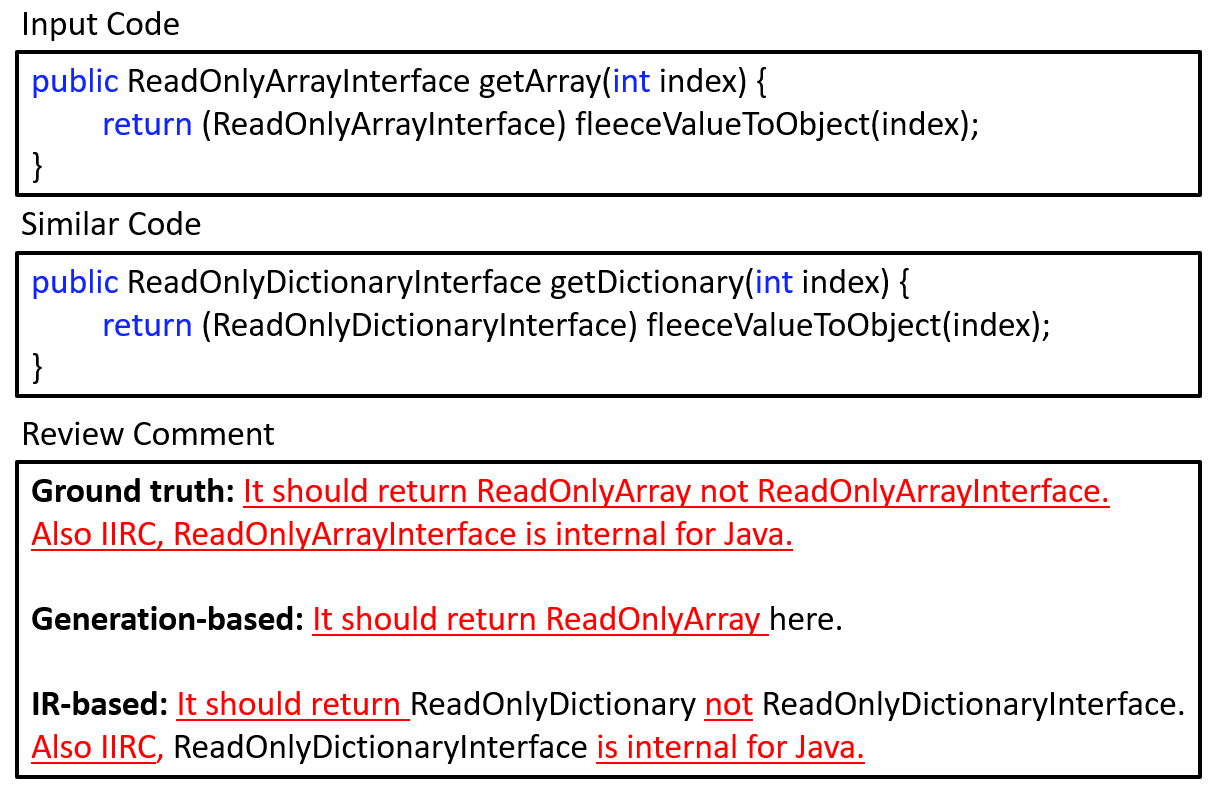}
\caption{A motivating example of generation-based and IR-based code review comment generation. Words in ground truth were underlined.}
\label{fig:motivating example}
\end{figure}
Low-Frequency Ground-Truth Tokens (LFGTs) \cite{arthur2016incorporating, zhang2018guiding}. Accurately generating LFGTs is critical for improving the overall quality and usefulness of generated review comments. We hypothesize that incorporating retrieval candidates from IR-based RCG can expose the generative model to informative exemplars containing LFGTs, thereby guiding the language model to generate contextually appropriate LFGTs.

Figure~\ref{fig:motivating example} shows a motivating example. This is an example result of generation-based RCG (CodeT5 \cite{wang2021codet5}) and IR-based (CommentFinder \cite{hong2022commentfinder}) RCG on the Tufano et al.’s benchmark dataset \cite{tufano2022using}. Given an input code snippet, the generation-based approach produces review comments using its parametric knowledge acquired during training. In contrast, the IR-based method identifies lexically similar code from the training dataset and directly reuses the associated review comment as the output. The generation-based model tends to produce high-frequency tokens such as “it” (46,654 occurrences in the training corpus), “should” (26,721), and “return” (142,643), but it fails to generate LFGTs like “Also” (7,476), “IIRC” (123), “internal” (1,621), and “Java” (7,016). Although the model successfully generate “ReadOnlyArray” (1), a token that appears only once in the training corpus, this is primarily due to its presence in the input code itself, rather than the model’s ability to generalize to rare vocabulary. Conversely, the IR-based method is capable of retrieving review comments that contain LFGTs such as “Also,” “IIRC,” “internal,” and “Java” by leveraging similar examples in the retrieval set. However, it fails to retrieve identifier-specific tokens like “ReadOnlyArray” when such tokens are rare in the retrieval corpus. This example illustrates a complementary gap between generation-based RCG and IR-based RCG. While generation-based RCG can effectively generate LFGTs that are present in the input code (e.g., “ReadOnlyArray”), it struggles to produce LFGTs that are absent from the input. Conversely, IR-based RCG can retrieve LFGTs from similar examples in the retrieval corpus even when they are not present in the input code (e.g., “IIRC” or “Java”), but it struggles to capture identifier tokens that appear in the input but are rare in the retrieval set. This observation motivates our investigation into a hybrid approach that integrates both paradigms to leverage their respective strengths.

To bridge the complementary gap between generation-based and IR-based RCG, we propose to leverage Retrieval Augmented Generation (RAG) \cite{lewis2020retrieval} to RCG by augmenting input code query with top-k most similar code’s review exemplars. RAG enhances generative language model’s performance on knowledge-intensive tasks by augmenting input queries with relevant knowledge passages \cite{lewis2020retrieval}. By incorporating retrieved exemplars, RAG enables generative language models to access domain-specific knowledge and refer to LFGTs without requiring additional training \cite{lu2024improving}. Building on this principle, RAG has been effectively employed in diverse software engineering tasks, including code summarization \cite{zhang2020retrieval}, code completion \cite{lu2022reacc}, code generation \cite{parvez2021retrieval}, and test case generation \cite{shin2024retrieval}. Given that code review requires extensive knowledge of project-specific practices—such as coding conventions, review focus areas, and stylistic guidelines \cite{han2021understanding}—RAG can be beneficial, as retrieved exemplars help convey this domain-specific knowledge to the generation model. However, the use of RAG for generating review comments has not been explored. To address this gap, we propose RAG-Reviewer: RAG-based RCG framework that integrates IR-based and generation-based approaches.

The RAG-Reviewer framework comprises two primary components: a retrieval module and a generation module. The retrieval module retrieves top-k similar review exemplars that are most similar to the input code snippet from a large historical code review database. The generation module utilizes a PLM to generate a review comments for the input code snippet with the retrieved exemplars. Given an input code for review, the retrieval module first identifies the top-k most similar code examples and extracts their corresponding review comments. These retrieved exemplars are then concatenated with the input code and passed to the generation module. To adapt the generator for this retrieval-augmented input format, we fine-tune the language model to produce review comments that jointly reflect information from both the input and the retrieved examples. Empirical evaluations demonstrate that RAG-Reviewer outperforms both generation-only and IR-only baselines. It achieves consistent improvements in Exact Match and BLEU scores across all PLMs and enhances the generation of LFGTs. The key contributions of this work are as follows:
\begin{itemize}
    \item We propose \textbf{RAG-Reviewer}, a RAG framework for RCG that unifies generation-based and IR-based approaches. To the best of our knowledge, this is the first study to apply the RAG paradigm to the RCG task.
    \item We conduct comprehensive experiments across multiple PLMs, showing that RAG-Reviewer improves generation accuracy, better captures LFGTs, and exhibits performance improvements as the number of retrieved exemplars increases.
    \item To support reproducibility and encourage future research, we publicly release our implementation on GitHub: 
    
    \url{https://github.com/RAG-Reviewer/RAG-Reviewer}.
\end{itemize}

\begin{figure}[t]  % Top of left column
\centering
\includegraphics[width=0.75\linewidth]{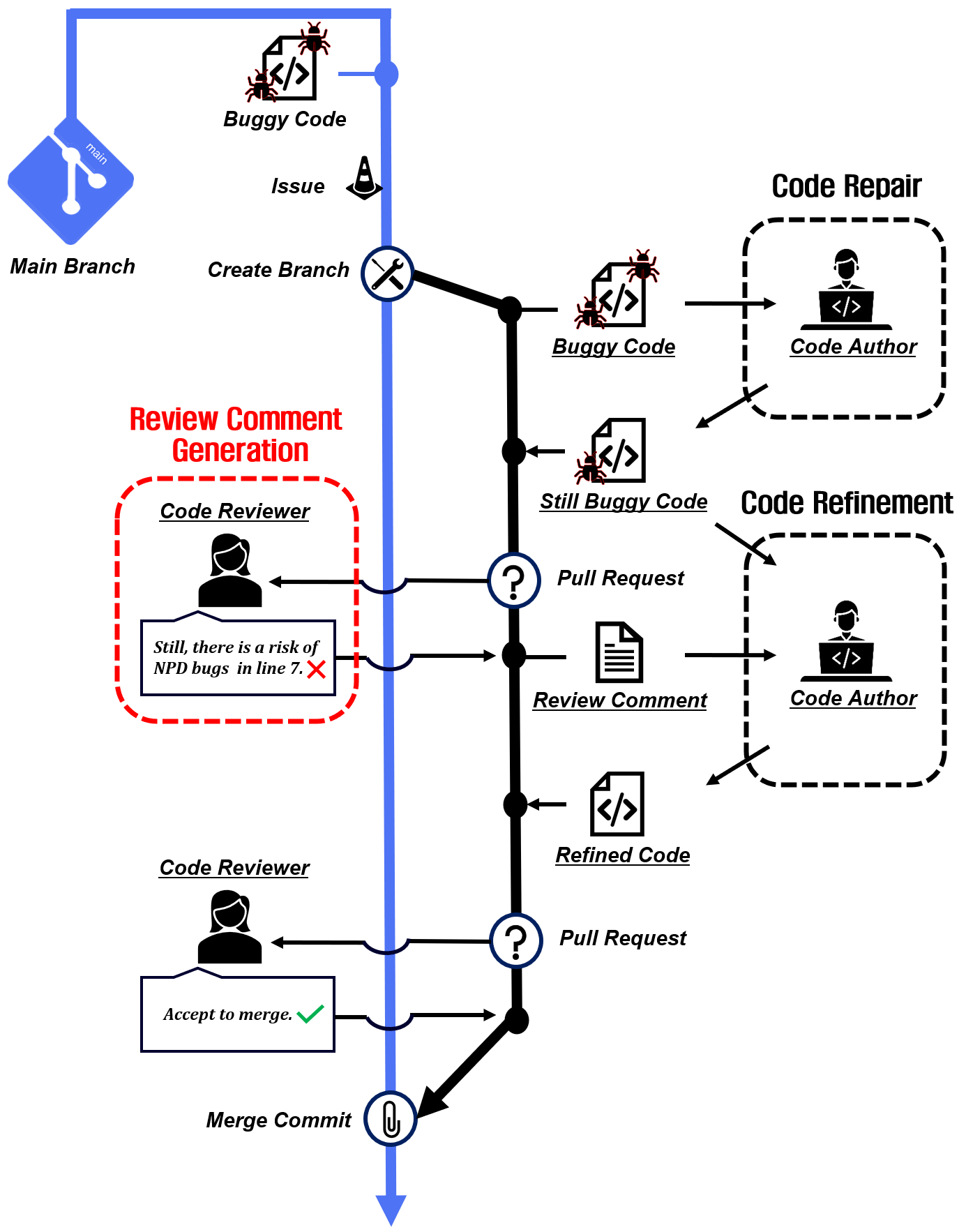}
\caption{Code review process workflow}
\label{fig:code_review_process}
\end{figure}

\section{Background}\label{sec:Background}
In this section, we provide a brief overview of code review process and code review comment generation.
\subsection{Code Review Process}\label{AA}
Code review is the one of the widely adapted software quality assurance practices in both open-source projects and industries~\cite{alami2019does, sadowski2018modern}. Through code review process, various aspects of code quality (e.g., readability, maintainability, testability, scalability, secure and more) can be inspected by collective expertise of reviewers. Figure~\ref{fig:code_review_process} shows code review process scenario in debugging the faulty code. Suppose that an issue occurred in main branch due to a bug. Then, (1) the author who is responsible for this issue repairs the buggy code and requests the code reviewer to pull repaired code into main branch. Later, (2) the code reviewer evaluates the proposed code to determine whether the revised code is suitable for integration into the main branch. If the code still contains defects, the reviewer rejects the pull request and provides feedback through a review comment. (3) The author subsequently revises the code based on the reviewer's suggestions and resubmits the pull request. (4) Upon re-evaluation, if the changes address the identified issues, the reviewer approves the code for merging into the main branch. Through this code review process, all codes can be inspected by peer reviewers before it integrated to the main codebase, thereby enhancing overall software quality. Within this process, code review process involves three tasks: code repair (code-to-code), review comment generation (code-to-comment) and code refinement (code\&comment-to-code) as illustrated in Figure~\ref{fig:code_review_process}. The field of Code Review Automation (CRA) aims to automate these three tasks~\cite{tufano2022using} and we focus on automated review comment generation in this work. 

\begin{figure}[t]  % Top of left column
\centering
\includegraphics[width=0.8\linewidth]{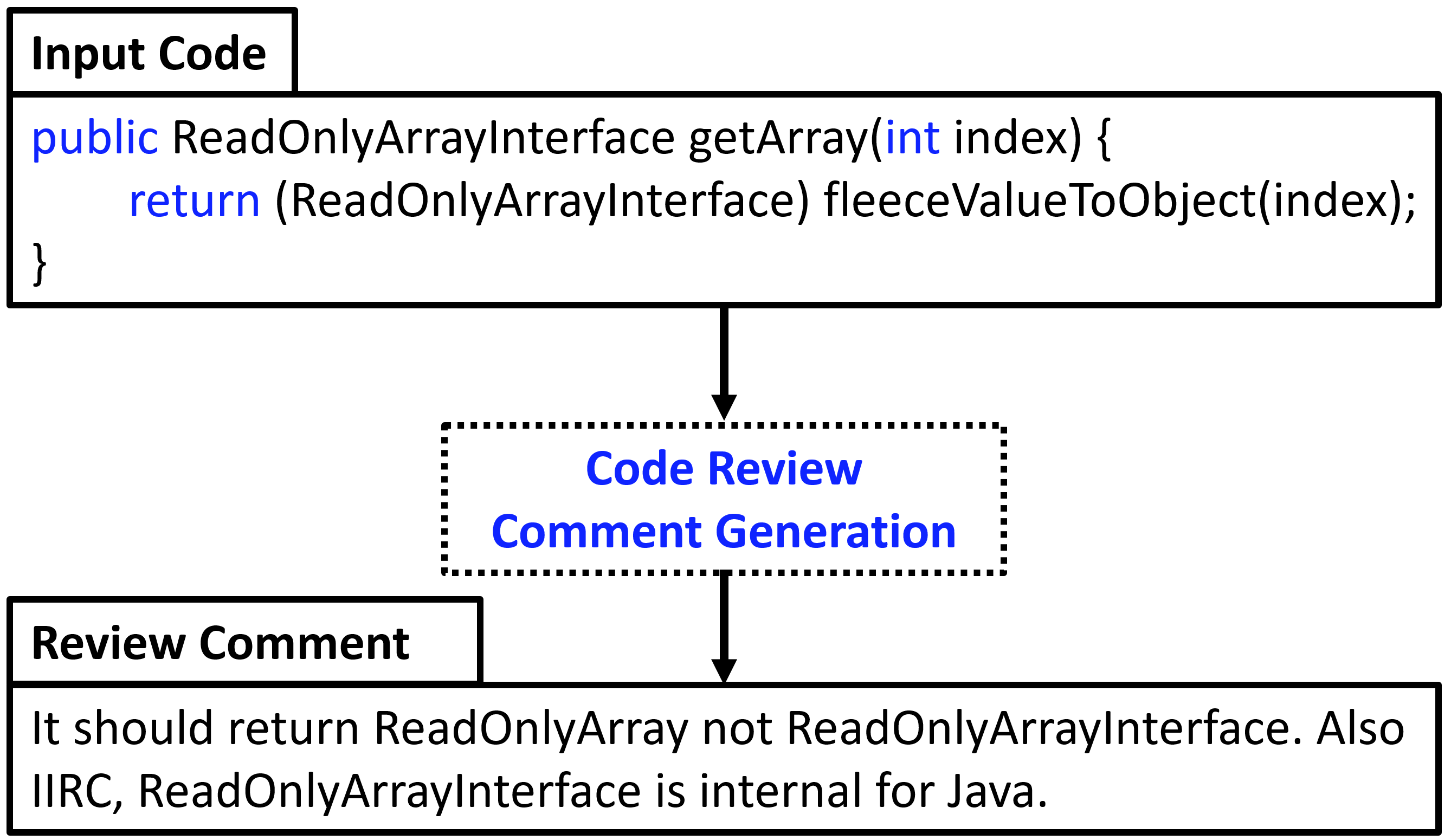}
\caption{An example of code review comment generation}
\label{fig:An example of code review comment generation}
\end{figure}

\subsection{Code Review Comment Generation}
Code Review Comment Generation (RCG) automates the generation of review comments in the code review process, as shown in Figure~\ref{fig:An example of code review comment generation}. RCG assists code reviewers by automatically generating review comments, which can serve as references or starting points for writing their own final feedback. Formally, given an input code sequence to be reviewed, $C_{input}=\{c_1,c_2,\dots,c_n\}$, and its corresponding code review comments, $N_{Review}=\{n_1,n_2,\dots,n_m\}$, generation-based RCG is a sequence-to-sequence probabilistic mapping function $G: C_{input} \rightarrow N_{Review}$. In a probabilistic sequence modeling framework, the probability of generating $N_{Review}$ given $C_{input}$ is defined as:\[
P_G(N_{Review} \mid C_{input}) = \prod_{i=1}^m P_G(n_i \mid C_{input}, n_1, \dots, n_{i-1})
\]where $P_G$ denotes a probabilistic autoregressive model parameterized by $G$.

\section{Approach} \label{sec:Approach}

\begin{figure*}[t]  % Top of the page, spanning both columns
\centering
\includegraphics[width=1\textwidth]{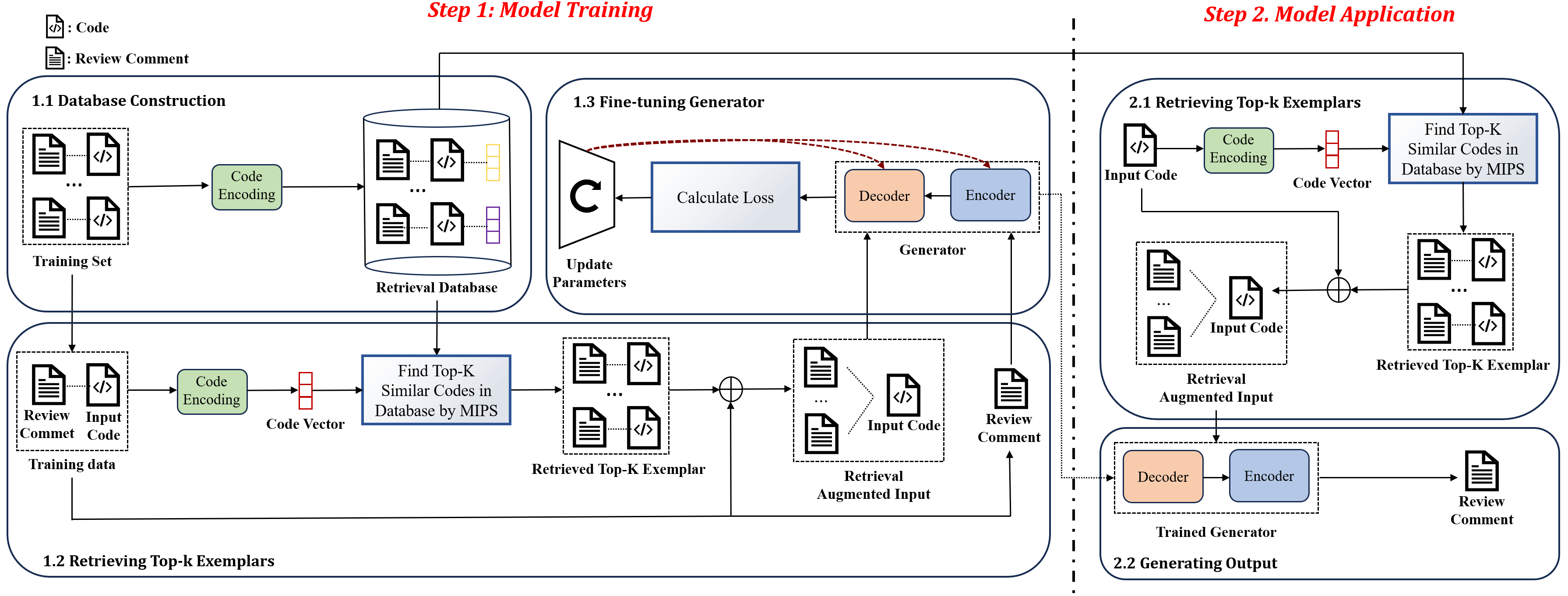}
\caption{Overall framework of RAG-Reviewer}
\label{fig:Overall framework of RAG-Reviewer}
\end{figure*}

\subsection{Overall Approach}
In this work, we propose \textbf{RAG-Reviewer}, a \textbf{RAG}-based code \textbf{Review} comment gen\textbf{er}ation framework. RAG-Reviewer integrates the strengths of both IR-based and generation-based methods. It consists of two core components: 
(a) \textbf{Retrieval Module} that retrieves review comments corresponding to code snippets that are most similar to the given input code and (b) \textbf{Generator Module} that produces review comments for the input code by conditioning on both the input and the retrieved exemplars.

By augmenting the input code with exemplars retrieved via the retrieval module, the generative language model can leverage the outputs of the IR-based approach to enhance the quality of generated review comments. Figure~\ref{fig:Overall framework of RAG-Reviewer} provides an overview of RAG-Reviewer, which operates in two phases: model training and model application. During the training phase, RAG-Reviewer constructs a retrieval database from the training dataset, comprising code snippets, their corresponding review comments, and their encoded vector representations. Each code snippet is encoded into a dense vector space to facilitate similarity computation using Maximum Inner Product Search (MIPS). The retrieval module then identifies the top-k most similar code snippets from this database using the same encoding mechanism. To prevent data leakage, the top-1 retrieved exemplar—which is identical to the input training instance—is excluded during training. The generator is fine-tuned on these augmented inputs, learning to condition on both the input code and the retrieved exemplars. During model application, the same retrieval procedure is applied to obtain relevant exemplars for a new input code snippet.
\subsection{Retrieval Module}
The retrieval module is responsible for identifying the most relevant exemplars from the retrieval database given an input code snippet. These retrieved exemplars provide contextual guidance that the generator can leverage to produce more accurate and informative review comments. The retrieval database is constructed from the training dataset. Formally, the training dataset is defined as a set $T_\text{review} = \{(IC_i, CR_i) | i=1,\dots,M\}$ where $IC_i$ denotes the input code snippet, $CR_i$ represents its corresponding ground-truth review comment and $M$ is the total number of training instances. This dataset serves as the basis for both retrieval and generator training within the RAG-Reviewer framework.
To enable similarity computation between the input code and the code snippets in the retrieval database, all input code instances $IC_i$s must be encoded into dense vector representations. This vectorization allows for the efficient measurement of semantic similarity—typically using cosine similarity—between code snippets. Formally, we employ an encoder  $d_c$:
\begin{equation}
h_i = d_c(IC_i), \forall (IC_i, CR_i) \in T_\text{review}
\end{equation}
where $h_i$ is the vectorized representation of $IC_i$. We use the UniXcoder~\cite{guo2022unixcoder} as $d_c$, because it can effectively preserve code’s semantic and syntactic information. Especially, UniXcoder showed state-of-the-art performance in IR-based RCG~\cite{kartal2024automating}. Based on this, we construct the retrieval database $D_\text{review}$ as follows:
\begin{table*}[!t]
\centering
\caption{Frequency distribution of review comments in Tufano et al.'s training dataset}
\label{tab:token_freq_distribution}
\resizebox{\textwidth}{!}{%
\begin{tabular}{c c c c c c c c}
\toprule
\textbf{Token Freq. ($\leq x$)} & $\leq 1$ & $\leq 5$ & $\leq 10$ & $\leq 20$ & $\leq 50$ & $\leq 100$ & All ($\geq 1$) \\
\midrule
\textbf{Unique Tokens Count} & 2,592 (10.08\%) & 8,827 (34.23\%) & 12,817 (49.84\%) & 16,558 (64.39\%) & 20,544 (79.88\%) & 22,507 (87.52\%) & 25,717 (100\%) \\
\textbf{\# Review Comments Containing Tokens} & 2,486 (1.85\%) & 16,944 (12.62\%) & 32,768 (24.41\%) & 52,597 (39.18\%) & 80,620 (60.06\%) & 98,550 (73.41\%) & 134,239 (100\%) \\
\bottomrule
\end{tabular}%
}
\end{table*}
\begin{table}[t]
\centering
\caption{Statistics of Tufano et al.’s dataset}
\label{tab:tufano_stats}
\resizebox{0.8\columnwidth}{!}{%
\begin{tabular}{lccc}
\toprule
\textbf{Dataset} & \textbf{Train} & \textbf{Valid} & \textbf{Test} \\
\midrule
\textbf{Count} & 134,239 & 16,780 & 16,780 \\
\textbf{Avg. Code Length} & 167 & 168 & 171 \\
\textbf{Avg. Review Length} & 26 & 26 & 26 \\
\bottomrule
\end{tabular}%
}
\end{table}
\begin{equation}
D_\text{review} = \{(h_i, IC_i, CR_i) | i = 1, \dots, M\}
\end{equation}
To find the most similar exemplars corresponding to a given input code $C_\text{input}$, we encode $C_\text{input}$ into $h_x$ using $d_c$:
\begin{equation}
h_x = d_c(C_\text{input})
\end{equation} 
Note that we used same encoder $d_c$ for both $C_\text{input}$ and $IC_i$ because they are both forms of code, making them unimodal. Using the same encoder ensures consistency in the dense vector representation. After encoding, we compute similarity score $s_i$ between $C_\text{input}$ and $IC_i$s using the inner dot product:
\begin{equation}
s_i = \langle h_x, h_i \rangle, \forall i \in \{1,\dots,M\}
\end{equation}
where $\langle \cdot, \cdot \rangle$ denotes the inner product. Based on the computed similarity scores, we then select the top-k most similar exemplars with respect to the input code:
\begin{equation}
\scalebox{0.80}{$
\mathcal{C}_R = \text{TopK}\left(\left\{s_i\right\}_{i=1}^M, \mathcal{D}_{\text{review}}, K\right) = \left\{\left(IC_1, CR_1\right), \ldots, \left(IC_k, CR_k\right)\right\}
$}
\end{equation}
where $\text{TopK}(S, D, K)$ returns top-k most similar code review exemplars based on the similarity score set $S$, retrieval database $D$ and specified value of $K$. The resulting top-k retrieval candidates $C_R$ are then used to augment the input in the generation module.
\subsection{Generation Module}
Given the output (Equation 5) of retrieval module, the list of top-k most similar exemplars with respect to the input code, we construct the input to the generator by concatenating the input code sequence $X$ with the retrieved candidates $C_R$:
\begin{equation}
X' = X \oplus C_1 \oplus C_2 \dots \oplus C_k,\quad C_i \in C_R
\end{equation} where $C_i$ denotes $i^{\text{th}}$ most similar code’s exemplar. Each $C_i$ consists of a code snippet $IC_i$ and its associated review comment $CR_i$. Thus, we have to decide what to retrieve for each $C_i$. Following methodology of Parvez et al.~\cite{parvez2021retrieval}, we explore two input augmentation strategies, \textbf{pairs} and \textbf{singleton}. The pair setting incorporates both the exemplar code and its corresponding review comment, whereas the singleton setting leverages only the review comment:

{
\begin{align}
X_p' &= X \oplus CR_1 \oplus IC_1 \cdots, \quad (IC_i, CR_i) \in \mathcal{C}_R \label{eq:pair_aug} \\
X_s' &= X \oplus CR_1 \oplus CR_2 \cdots, (IC_i, CR_i) \in \mathcal{C}_R  \label{eq:singleton_aug}
\end{align}
}where $IC_i$ means $i^{\text{th}}$ most similar code and $CR_i$ means its corresponding review comment retrieved from database $D_\text{review}$. To clearly delineate the boundaries between code and review comments within each exemplar, we introduce special delimiter tokens, [csep] and [nsep], following the input formatting strategy proposed by Parvez et al.~\cite{parvez2021retrieval}.

Given the augmented input $X'$, RAG-Reviewer generate the output sequence $Y = (y_1,\dots,y_N)$ using a generative language model under an autoregressive formulation:
\begin{equation}
P_\theta(Y \mid X') = \prod_{i=1}^N P_\theta(y_i \mid X', y_{1:i-1})
\end{equation}
where $P_\theta$ is a language model parameterized by $\theta$, which generates each token conditioned on the input $X'$ and previous $i-1$ tokens. We experiment with various types of pre-trained language models as the generator and compare their performance within the RAG-Reviewer framework.

To train the generator, RAG-Reviewer minimizes the following negative log-likelihood loss function over the target sequence $Y^\text{true} = (y_1^\text{true}, \dots, y_N^\text{true})$:

\begin{equation}
L(\theta) = -\sum_{i=1}^N \log P_\theta(y_i^\text{true} \mid X', y_{1:i-1}^\text{true}) \tag{10}
\end{equation}
where $\theta$ denotes the trainable parameters of the generator.
We fine-tuned only the generator while keeping the code encoder $d_c$ fixed because training $d_c$ would necessitate updating $h_i$ (Equation 1) at every gradient step, as changes to $d_c$ would alter $h_i$, making recalculations computationally expensive~\cite{lewis2020retrieval}.

\section{Experiment Setup} \label{sec:Experiment Setup}

\subsection{Research Question}
We formulate the following research questions to evaluate the effectiveness of RAG-Reviewer:
\begin{itemize}
    \item \textbf{RQ1:} Does RAG-Reviewer outperform existing baseline methods in terms of review comment generation quality?
    \item \textbf{RQ2:} Does RAG-Reviewer mitigate the challenges faced by language models in generating low-frequency tokens?
    \item \textbf{RQ3:} What is the impact of the number of retrieved exemplars on the performance of RAG-Reviewer?
\end{itemize}
\begin{figure}[t]
\hfill
\begin{minipage}{0.48\textwidth}
    \centering
    \begin{minipage}{0.48\textwidth}
        \includegraphics[width=\linewidth]{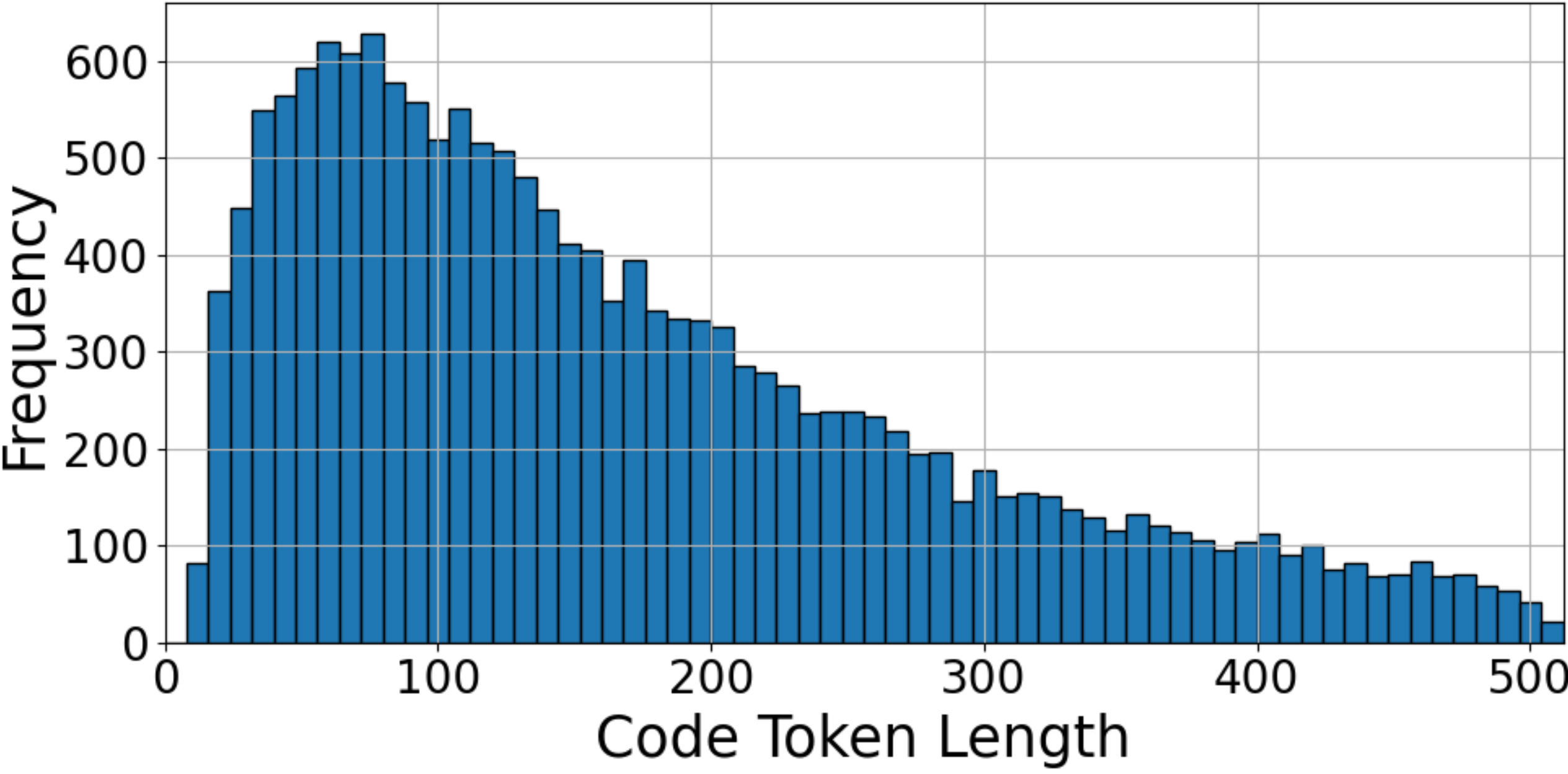}
        \centering
        {\small (a) Code length distribution}
    \end{minipage}
    \hfill
    \begin{minipage}{0.48\textwidth}
        \includegraphics[width=\linewidth]{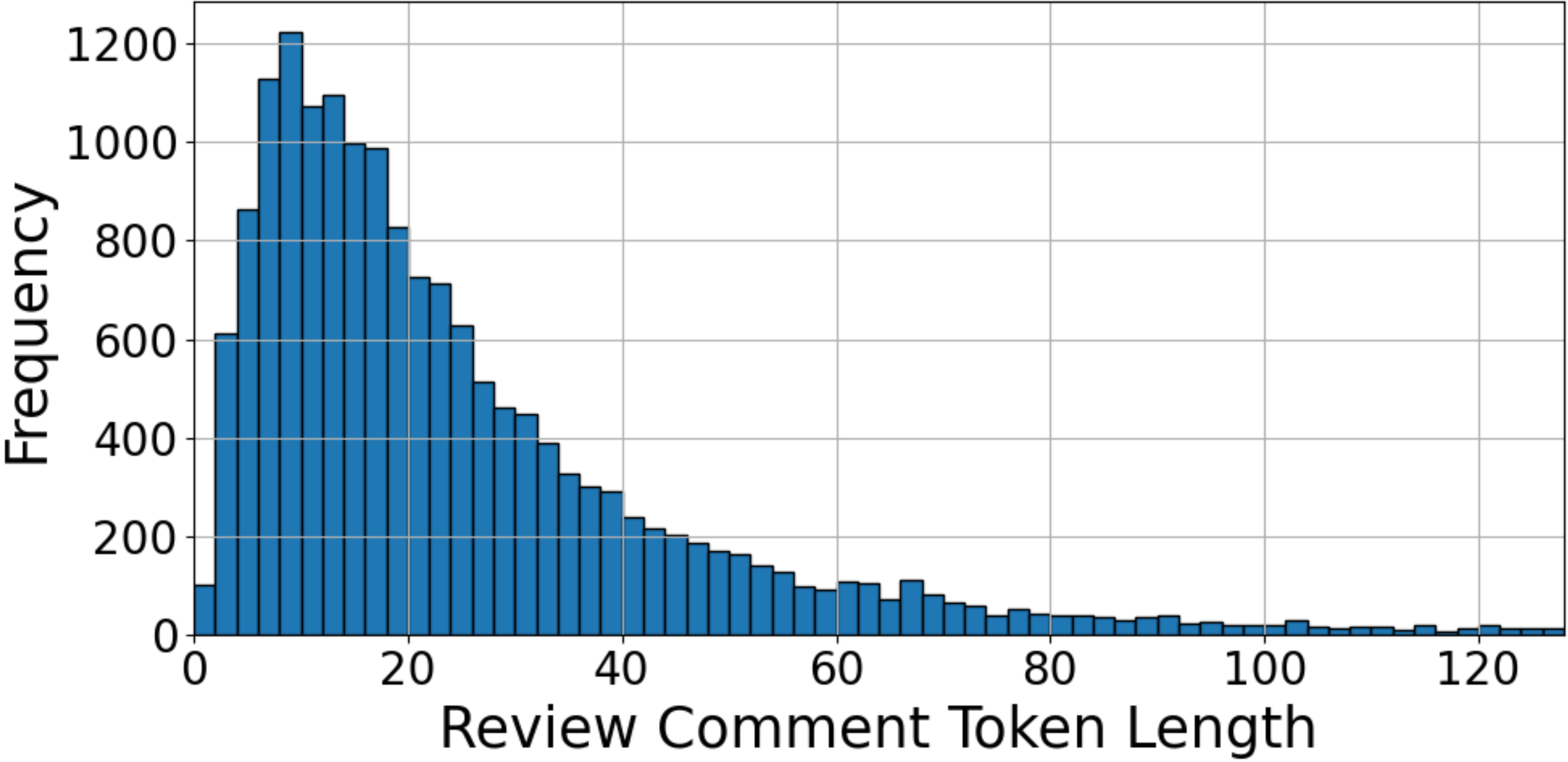}
        \centering
        {\small (b) Review length distribution}
    \end{minipage}
    \vspace{2mm}

    \caption{\small Token length distribution of Tufano et al.'s test dataset}
    \label{fig:token_length_distribution}
\end{minipage}
\hfill
\end{figure}
These research questions are empirically investigated in Section~\ref{sec:Experimental Result}.
\subsection{Dataset}
We use the dataset introduced by Tufano et al.~\cite{tufano2022using}. which was constructed from large-scale Java open-source projects on GitHub and Gerrit. Each code is function-level granularity written in Java and paired with its corresponding review comment. Detailed statistics are shown in Table~\ref{tab:tufano_stats}, and Figure~\ref{fig:token_length_distribution} illustrates the token length distribution in the test set. All lengths were calculated using the CodeT5 ~\cite{wang2021codet5} tokenizer. As shown in Table~\ref{tab:token_freq_distribution}, 87.52\% of unique tokens in the review comment training corpus occur no more than 100 times, and 73.41\% of comments contain at least one such token. Following prior work~\cite{zhang2020retrieval, wei2020retrieve}, we define these as low-frequency tokens. This high prevalence makes the Tufano et al. dataset a suitable benchmark for evaluating our approach to improving low-frequency token generation.

\subsection{Evaluation Metrics}
Following the previous code review comment generation work~\cite{tufano2022using, hong2022commentfinder, li2022codereviewer, zhou2023generation}, we evaluate the quality of generated review comments using the metric BLEU~\cite{papineni2002bleu} and Exact Match (EM) score. These metrics are widely used in other automated software engineering work~\cite{zhang2020retrieval, parvez2021retrieval, lu2022reacc, wei2020retrieve}.

\textbf{BLEU} measures the frequency of overlapped n-grams between the generated comment and the reference comment using n-gram precision. We use BLEU-4, which considers up to 4-gram matches. The BLEU-N score is computed as:
\begin{equation*}
\text{BLEU-N} = \text{BP} \cdot \exp\left(\sum_{n=1}^{N} w_n \log p_n \right)
\end{equation*}
where $p_n$ is the n-gram precision, $w_n$ is the weight for each n-gram (uniformly set to $1/N$), and $\text{BP}$ is a brevity penalty that penalizes overly short generated sequences. Higher BLEU scores indicate greater similarity to the ground truth review comment, ranging from 0\% to 100\%.

\textbf{EM} measures the ratio of generated outputs that are exactly identical to the ground truth review comment. It is a stricter metric than BLEU because it requires a character-wise match. The EM score also ranges from 0\% to 100\%, where 100\% indicates that all the generated outputs for the test dataset are character-wise identical to the ground truth.

\subsection{Baselines}
To evaluate the effectiveness of \textbf{RAG-Reviewer}, we compare it against a range of baselines, including both generation-based and IR-based methods. We reproduced all the baselines.

\begin{itemize}
    \item \textbf{CommentFinder}: A first IR-based RCG approach that uses Bag-of-Words (sparse vector) for code vectorization and cosine similarity for candidate ranking~\cite{hong2022commentfinder}.
    
    \item \textbf{UniXCoder-IR}: Proposed by Kartal et al.~\cite{kartal2024automating}, this method uses UniXCoder (dense vector)~\cite{guo2022unixcoder} for code vectorization. Retrieval is based on similarity metrics such as cosine or Euclidean distance.
    
    \item \textbf{Tufano T5}: A small version of the T5~\cite{raffel2020exploring} architecture (6-layer encoder and 6-layer decoder) pre-trained for code review automation. It was pre-trained on code and natural language pairs collected from Stack Overflow and CodeSearchNet~\cite{husain2019codesearchnet} using masked language modeling~\cite{tufano2022using}.
    
    \item \textbf{CodeT5}: A pre-trained encoder-decoder model that supports both code understanding and generation. Based on the T5~\cite{raffel2020exploring} architecture, CodeT5~\cite{wang2021codet5} introduces identifier-aware objectives and bimodal dual generation tasks. It was pre-trained on CodeSearchNet~\cite{husain2019codesearchnet} and additional GitHub repositories. We use the \texttt{CodeT5-base} variant, which has 12 encoder and 12 decoder layers.
    
    \item \textbf{CodeT5+}: An extended version of CodeT5 that incorporates a mixture of pre-training objectives including span denoising, contrastive learning, text–code matching, and causal language modeling. CodeT5+~\cite{wang2023codet5+} is trained on both unimodal code and bimodal code–text corpora. We use the \texttt{CodeT5p-220m} version.
    
    \item \textbf{CodeReviewer}: Specifically pre-trained for code review automation tasks, CodeReviewer~\cite{li2022codereviewer} leverages a multilingual dataset of code diffs and review comments from GitHub. It introduces four novel pre-training tasks tailored for code review, such as diff tag prediction and denoising objectives for both code and review comments.
    
    \item \textbf{AUGER}: AUGER~\cite{li2022auger} is a T5-based~\cite{raffel2020exploring} model that enhances review comment generation using review-line tagging and cross pre-training. It is trained on a curated dataset of over 10,000 real-world code change and review comment pairs from popular Java projects. AUGER incorporates review-specific tags and demonstrates significant improvements in review comment quality.
\end{itemize}

\subsection{Implementation Details}
All experiments were conducted on a NVIDIA Tesla V100-SXM2-32GB GPU. For \texttt{CodeT5, CodeT5+} and \texttt{CodeReviewer}, we used the AdamW optimizer (learning rate = 3e-5, weight decay = 0.01) with a linear learning rate schedule and 10\% warm-up steps. To address memory constraints, we applied gradient accumulation over 3 steps with a batch size of 12 (effective batch size = 36). We used gradient clipping with a maximum norm of 1.0 and enabled mixed-precision training. The input and output lengths were capped at 512 tokens for code and 128 tokens for review comments, based on the dataset's token distribution (see Table~\ref{tab:tufano_stats} and Figure~\ref{fig:token_length_distribution}). Longer input sequences were truncated to fit within token limits. For RAG-based models, input prompts were augmented with as many retrieved exemplars as would fit within the input token budget. We used a beam size of 10 for evaluation. Training was conducted for up to 20 epochs, with early stopping applied after 3 epochs of no improvement on the validation set. The best model checkpoint was selected based on validation performance. For \texttt{Tufano T5} and \texttt{AUGER}, we used the Adam optimizer (learning rate = 3e-4) with a batch size of 12 and no gradient accumulation, based on preliminary tuning results. All baseline models were implemented using publicly available replication packages~\cite{tufano2021replication, codet5replication, codet5preplication, codereviewerreplication, augereplication, commentfinderreplication, unixcoderreplication}.

\section{Experimental Result} \label{sec:Experimental Result}
In this section, we present experimental result of our work following three research questions.
\subsection{\textbf{(RQ1) Does RAG-Reviewer outperform existing baseline methods in terms of review comment generation quality?}}

\textbf{Singleton vs. Pair Retrieval Strategy.} Before comparing RAG-Reviewer with existing baselines, we first evaluate which retrieval strategy achieves superior performance within RAG-Reviewer. Following Parvez et al.~\cite{parvez2021retrieval}, we consider two strategies: (1) \textit{Singleton Retrieval}, which retrieves a larger number of review comments without their associated code snippets and (2) \textit{Pair Retrieval}, which retrieves fewer exemplars but preserves both the code snippets and their corresponding review comments.

\begin{table}[t]
\centering
\caption{Performance comparison of RAG-Reviewer with generation-based and IR-based baselines}
\label{tab:performance_comparison}
{\fontsize{8pt}{9pt}\selectfont
\resizebox{0.75\columnwidth}{!}{%
\begin{tabular}{lcc}
\toprule
\textbf{Name} & \textbf{EM (\%)} & \textbf{BLEU (\%)} \\
\midrule
\multicolumn{3}{l}{\textbf{Information Retrieval}} \\
CommentFinder & 2.80 & 12.41 \\
UniXCoder-IR & 2.79 & 12.80 \\
\midrule
\multicolumn{3}{l}{\textbf{Generation}} \\
Tufano T5 & 0.87 & 9.32 \\
CodeReviewer & 1.23 & 9.27 \\
CodeT5 & 1.39 & 9.91 \\
CodeT5+ & 1.71 & 11.59 \\
AUGER & 1.10 & 9.78 \\
\midrule
\multicolumn{3}{l}{\textbf{Retrieval Augmented Generation (Singleton)}} \\
Singleton Tufano T5 & 2.01 & 11.82 \\
Singleton CodeReviewer & 2.32 & 11.76 \\
Singleton CodeT5 & 2.40 & 12.45 \\
Singleton CodeT5+ & 2.53 & 11.97 \\
Singleton AUGER & 2.32 & 11.81 \\
\midrule
\multicolumn{3}{l}{\textbf{Retrieval Augmented Generation (Pair)}} \\
Pair Tufano T5 & 2.54 & 12.54 \\
Pair CodeReviewer & 2.76 & \textbf{13.52} \\
Pair CodeT5 & 2.90 & 12.98 \\
Pair CodeT5+ & \textbf{3.01} & 12.39 \\
Pair AUGER & 2.46 & 12.93 \\
\bottomrule
\end{tabular}%
}
}
\end{table}

As shown in Table~\ref{tab:performance_comparison}, the pair retrieval strategy consistently outperforms singleton retrieval in terms of both EM and BLEU scores across all PLMs. The EM gains range from +0.14\% (AUGER: 2.32\% $\rightarrow$ 2.46\%) to +0.53\% (Tufano T5: 2.01\% $\rightarrow$ 2.54\%). The BLEU gains range from +0.53\% (CodeT5: 12.45\% $\rightarrow$ 12.98\%) to +1.76\% (CodeReviewer: 11.76\% $\rightarrow$ 13.52\%). These results suggest that providing both code and its review comment in retrieved exemplars helps the model better learn code--comment relationships. Given the inherent input token limitations of language models, it is crucial to include helpful information—both in quantity and quality—within the available budget. Our experiments show that rather than allocating tokens to more review-only exemplars, it is more effective to include both the review comments and their corresponding code. Notably, since code snippets are typically much longer than comments (see Table~\ref{tab:tufano_stats}), pair retrieval fits much fewer exemplars within the token limit. Nevertheless, providing richer contextual information through paired code-comment exemplars yields better performance than using a larger number of comment-only examples.

\textbf{Generation-based vs. RAG-Reviewer.} RAG-Reviewer consistently outperforms generation-based models across all PLMs in both EM and BLEU scores. Notably, even models with modest baseline performance, such as Tufano T5 and CodeReviewer, show the most significant gains—Tufano T5 improves from 0.87\% to 2.54\% in EM (+1.67\%), while CodeReviewer improves from 9.27\% to 13.52\% in BLEU (+4.25\%). These results suggest that retrieval augmentation benefits not only strong backbones but also weaker ones, enhancing both precision and fluency by grounding generation in relevant contextual exemplars.

\textbf{IR-based vs. RAG-Reviewer.} Compared to IR-based methods, RAG-Reviewer achieves comparable or slightly higher performance in most cases. For example, Pair CodeT5 (2.90\%) and Pair CodeT5+ (3.01\%) slightly outperform CommentFinder (2.80\%) and UniXCoder-IR (2.79\%) in EM. BLEU scores show modest gains as well, with Pair CodeReviewer reaching 13.52\% versus 12.41\% and 12.80\% for the IR baselines. While the margins are small, RAG-Reviewer offers the added benefit of generating comments conditioned on both the input code and retrieved exemplars, allowing it to leverage semantic context that IR methods may overlook. Moreover, as shown in Table~\ref{tab:performance_comparison}, RAG-Reviewer’s performance generally improves with stronger PLMs, suggesting that its effectiveness scales with the capacity of the underlying generator.

\begin{tcolorbox}[colback=gray!5!white,colframe=black!75!black,title=Answering RQ1]
Pair retrieval consistently outperforms singleton retrieval in both EM and BLEU, and RAG-Reviewer achieves higher performance than generation-based and IR-based methods.
\end{tcolorbox}

\begin{table}[t]
\centering
\caption{The number of correctly generated low-frequency review tokens}
\label{tab:low_freq_token_generation}
\resizebox{\columnwidth}{!}{%
\begin{tabular}{lccccc}
\toprule
\textbf{Token Freq. ($\leq x$)} & \textbf{$\leq 20$} & \textbf{$\leq 40$} & \textbf{$\leq 60$} & \textbf{$\leq 80$} & \textbf{$\leq 100$} \\
\midrule
Vanilla Tufano T5 & 1102 & 1891 & 2450 & 2866 & 3238 \\
Pair Tufano T5 & 1342 & 2282 & 3030 & 3508 & 3917 \\
\% Improvement & \textbf{21.78\%} & \textbf{20.68\%} & \textbf{23.67\%} & \textbf{22.40\%} & \textbf{20.97\%} \\
\midrule
Vanilla CodeReviewer & 867 & 1525 & 2007 & 2405 & 2703 \\
Pair CodeReviewer & 1043 & 1869 & 2473 & 2936 & 3352 \\
\% Improvement & \textbf{20.30\%} & \textbf{22.56\%} & \textbf{23.22\%} & \textbf{22.08\%} & \textbf{24.01\%} \\
\midrule
Vanilla CodeT5 & 975 & 1770 & 2294 & 2711 & 3080 \\
Pair CodeT5 & 1018 & 1835 & 2429 & 2876 & 3247 \\
\% Improvement & \textbf{4.41\%} & \textbf{3.67\%} & \textbf{5.88\%} & \textbf{6.09\%} & \textbf{5.42\%} \\
\midrule
Vanilla CodeT5+ & 1142 & 1956 & 2557 & 3089 & 3489 \\
Pair CodeT5+ & 1229 & 2185 & 2909 & 3426 & 3856 \\
\% Improvement & \textbf{7.62\%} & \textbf{11.71\%} & \textbf{13.77\%} & \textbf{10.91\%} & \textbf{10.52\%} \\
\midrule
Vanilla AUGER & 807 & 1583 & 2232 & 2708 & 3073 \\
Pair AUGER & 879 & 1751 & 2509 & 3043 & 3472 \\
\% Improvement & \textbf{8.92\%} & \textbf{10.61\%} & \textbf{12.41\%} & \textbf{12.37\%} & \textbf{12.98\%} \\
\bottomrule
\end{tabular}%
}
\end{table}

\subsection{\textbf{(RQ2) Does RAG-Reviewer mitigate the challenges faced by language models in generating low-frequency tokens?}}

\textbf{Effect on Generating Low-frequency Tokens.} A well-known challenge for language models in generation-based methods is their difficulty in generating low-frequency ground-truth tokens (LFGTs), due to a bias toward high-frequency patterns learned during training \cite{arthur2016incorporating, zhang2018guiding}. In contrast, IR-based approaches can help expose such tokens by retrieving similar examples that contain them. While we have observed that RAG-Reviewer improves generation over standard baselines in RQ1, we further examine whether it specifically helps mitigate this limitation by evaluating its ability to generate LFGTs compared to generation-based methods. We define LFGTs as tokens that appear fewer than 100 times in the training review comment corpus following previous work \cite{zhang2020retrieval, wei2020retrieve}. For each generated review comment, we count a token as correctly generated if it appears in both the output and the ground truth. We then group these correctly generated tokens by their frequency in the training set using thresholds of $\leq$20, $\leq$40, $\leq$60, $\leq$80, and $\leq$100. The improvement is measured as the relative increase over the generation-based baseline.

As shown in Table~\ref{tab:low_freq_token_generation}, RAG-Reviewer improves the generation of LFGTs across all evaluated models. Tufano T5 exhibits the most consistent relative gains, ranging from 20.68\% to 23.67\%. And CodeReviewer shows similar improvements (ranging from 20.30\% to 24.01\%), while AUGER demonstrates increases between 8.92\% and 12.98\%. CodeT5+ shows improvements of 7.62\% to 13.77\%, and CodeT5 achieves more modest gains between 3.67\% and 6.09\%. These results suggest that retrieval augmentation can help generation models recover rarer tokens by grounding generation in retrieved exemplars. While the level of improvement varies by model, the overall result indicates that RAG-Reviewer helps language model generate LFGTs more effectively.

\begin{figure}[t]
    \centering
    \includegraphics[width=0.85\linewidth]{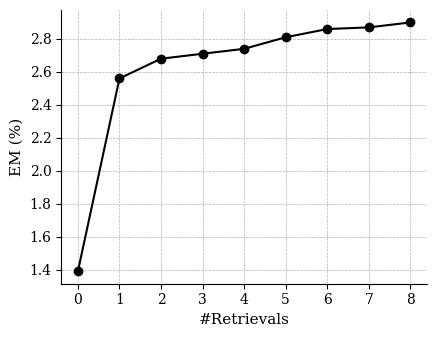}
    \caption{Exact match performance of Pair CodeT5 with varying numbers of retrieved exemplars on Tufano et al. benchmark}
    \label{fig:retrieval_ablation}
\end{figure}

\begin{tcolorbox}[colback=gray!5!white,colframe=black!75!black,title=Answering RQ2]
RAG-Reviewer enhances the generation of low-frequency tokens across all models, alleviating innate limitation of generation-based methods.
\end{tcolorbox}

\vspace{-0.9em}  % <--- Shrinks space after tcolorbox

\subsection{\textbf{(RQ3) What is the impact of the number of retrieved exemplars on the performance of RAG-Reviewer?}}

\textbf{Impact of Retrieval Quantity on Generation Quality.}  While RQ1 showed that exemplar quality (Pair) is more important than quantity (Singleton), we further investigate whether increasing the number of high-quality exemplars still improves performance. To explore this, we evaluated the Pair CodeT5 model using retrieval sizes ranging from 0 (vanilla) to 8. This setup was chosen based on findings from RQ1: pair retrieval consistently outperformed singleton retrieval across all backbones, and CodeT5 showed mid-range performance among the RAG-Reviewer variants—making it a representative case for analysis.

As shown in Figure~\ref{fig:retrieval_ablation}, exact match (EM) improves steadily from 1.39\% with no retrieval (vanilla CodeT5) to 2.90\% with eight exemplars. The biggest jump occurs when using just one exemplar (+1.17\%), suggesting that even a small amount of retrieval helps the model generate more accurate comments. Additional exemplars continue to improve performance, but the gains become smaller. This is likely due to the model’s input token limit, which restricts how much exemplar information can be included, thereby reducing the benefit of adding more examples.

\begin{tcolorbox}[colback=gray!5!white,colframe=black!75!black,title=Answering RQ3]
Including more retrieved exemplars in the input consistently improves RAG-Reviewer’s generation quality. However, performance gains diminish with higher counts due to input length constraints of the language model.
\end{tcolorbox}

\begin{figure}[t]
\centering

{
    \includegraphics[width=0.47\columnwidth]{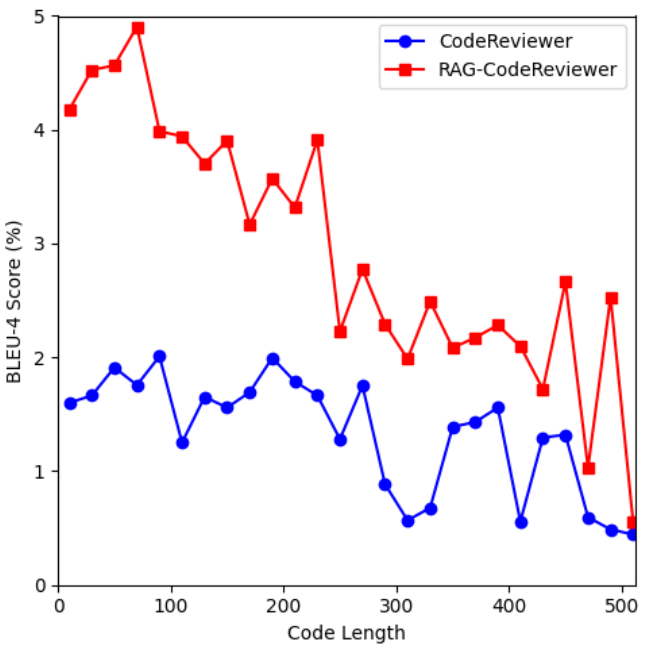}
}
\hfill
{
    \includegraphics[width=0.47\columnwidth]{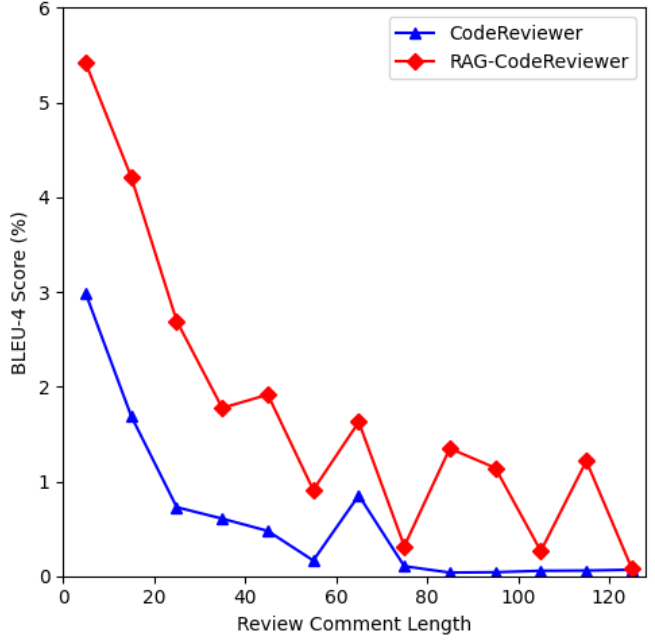}
}

\vspace{1em}

{
    \includegraphics[width=0.47\columnwidth]{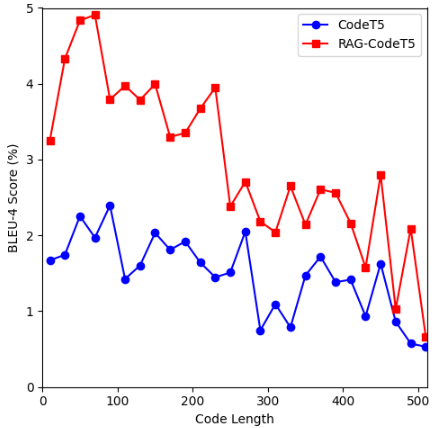}
}
\hfill
{
    \includegraphics[width=0.47\columnwidth]{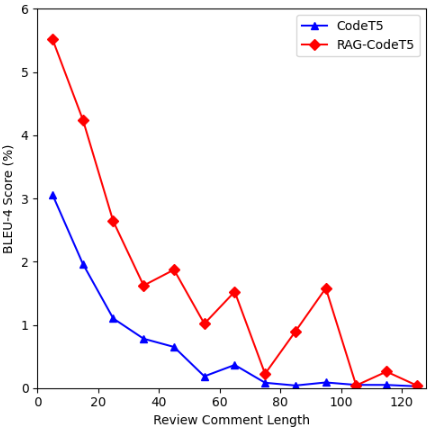}
}

\caption{BLEU-4 scores across different code and review comment lengths}
\label{fig:bleu_scores_lengths}
\end{figure}

\section{Discussion} \label{sec:Discussion}
This discussion covers three parts: how RAG-Reviewer performs with different code and review comment lengths, findings from manual analysis, and directions for future work.

\subsection{Performance with Different Length} 

To evaluate how RAG-Reviewer performs across varying input and output lengths, we compare BLEU-4 scores of Pair CodeReviewer and Pair CodeT5 with their vanilla counterparts. Test samples are bucketed by token length—20-token intervals for code and 10-token intervals for review comments—and average BLEU-4 is computed per bucket. As shown in Figure~\ref{fig:bleu_scores_lengths}, both RAG models consistently outperform their vanilla versions across all length intervals. Although BLEU-4 scores decrease with longer sequences, the RAG models maintain relatively higher performance.

\subsection{Manual Analysis}

\begin{table}[t]
\centering
\caption{Manual analysis of review comments on 100 input code from the Tufano et al. manual analysis samples}
\label{tab:manual_analysis}
{\fontsize{8.5pt}{9pt}\selectfont
\resizebox{\columnwidth}{!}{%
\begin{tabular}{lccc}
\toprule
\textbf{Category} & \textbf{RAG-Reviewer} & \textbf{CommentFinder} & \textbf{Tufano T5} \\
\midrule
Exact Match & 2 & 4 & 0 \\
Semantically Equivalent & 39 & 30 & 36 \\
Alternative Solution & 9 & 9 & 10 \\
Incorrect & 50 & 57 & 54 \\
\bottomrule
\end{tabular}%
}
}
\end{table}

\begin{table}[t]
\centering
\caption{Retrieved exemplars used for input augmentation in Figure~8 (Key LFGT `try-with-resources` underlined)}
\label{tab:retrieved_exemplars}
\resizebox{\columnwidth}{!}{%
\begin{tabular}{cl}
\toprule
\textbf{Top-K} & \textbf{Retrieved Review Comment} \\
\midrule
1 & STDERR as exit() follows \\
10 & Should we use \underline{try-with-resources} here to close it later? Test only so I think it may not matter much. \\
16 & This should be part of the \underline{try-with-resources} block. \\
20 & The \underline{try-with-resources} takes care of the closing, right? \\
\bottomrule
\end{tabular}
}
\end{table}

While RQ1 showed that RAG-Reviewer outperforms both IR-based and generation-based methods in terms of EM and BLUE, these metrics does not fully reflect the quality of generated review comments. To evaluate outputs beyond these metrics, we conducted a manual analysis of 100 review comments generated by the best-performing RAG-Reviewer (Pair CodeT5+), using the same 100 samples manually analyzed in Tufano et al. \cite{tufano2022using}, following the human evaluation method of prior work \cite{tufano2022using, hong2022commentfinder}. Each generated review comment was classified into one of four categories based on the labeling scheme introduced in Tufano et al. \cite{tufano2022using}: 
(a) \textit{Exact Match} – the generated review comment is character-wise identical to the ground-truth comment; 
(b) \textit{Semantically Equivalent} – the generated review comment differs in wording but conveys the same meaning as the ground truth’s intention; 
(c) \textit{Alternative Solution} – the generated review comment is neither identical nor semantically equivalent but still provides a meaningful and useful suggestion; 
(d) \textit{Incorrect} – the generated review comment is irrelevant and unhelpful for the given input code. 

\begin{figure}[t]
    \centering
    \includegraphics[width=1\columnwidth]{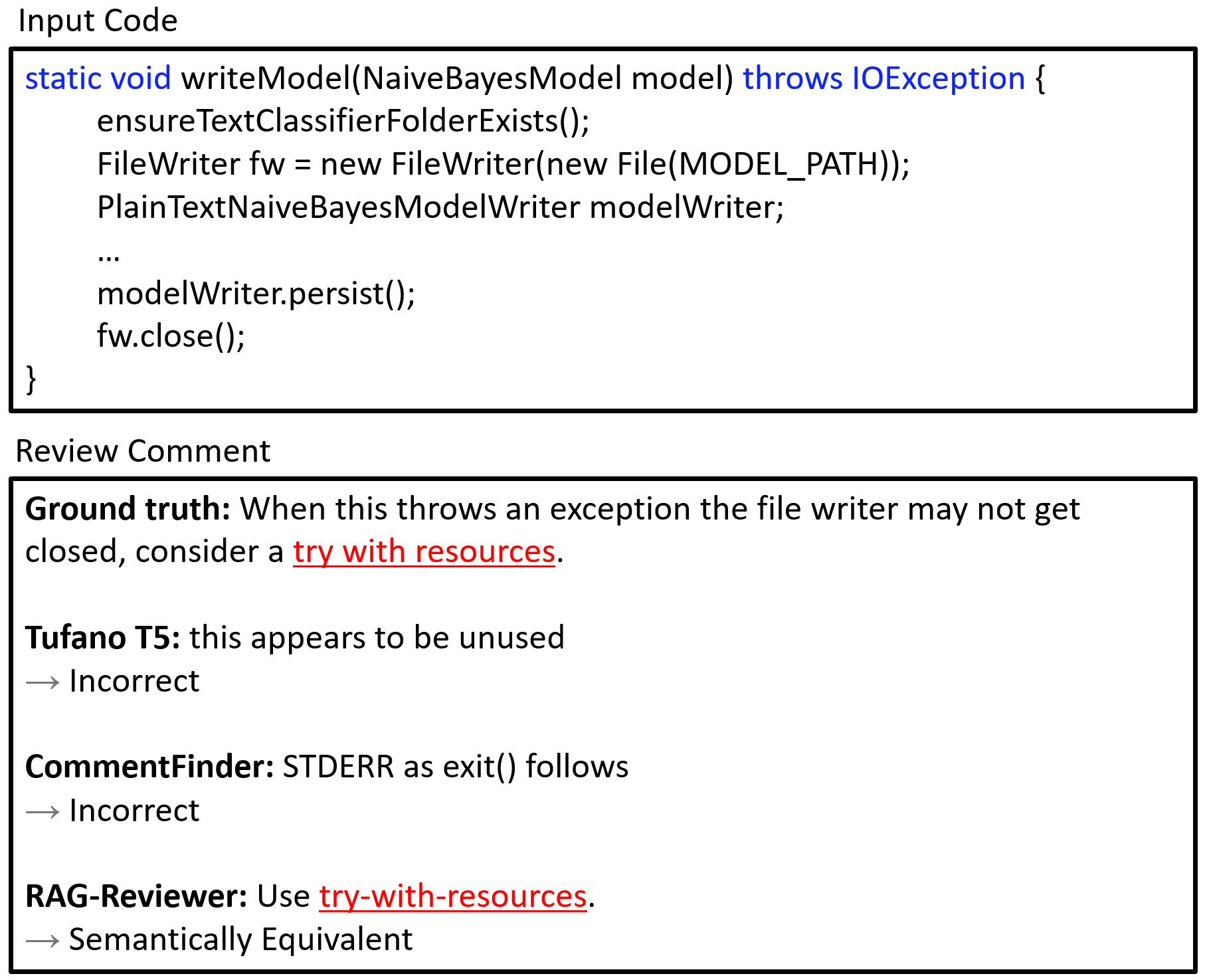}
    \caption{Example of a Semantically Equivalent review comment from the Tufano et al. manual analysis samples (Key LFGT `try-with-resources` underlined)}
    \label{fig:manual_example_semantic}
\end{figure}

\begin{figure}[t]
    \centering
    \includegraphics[width=1\columnwidth]{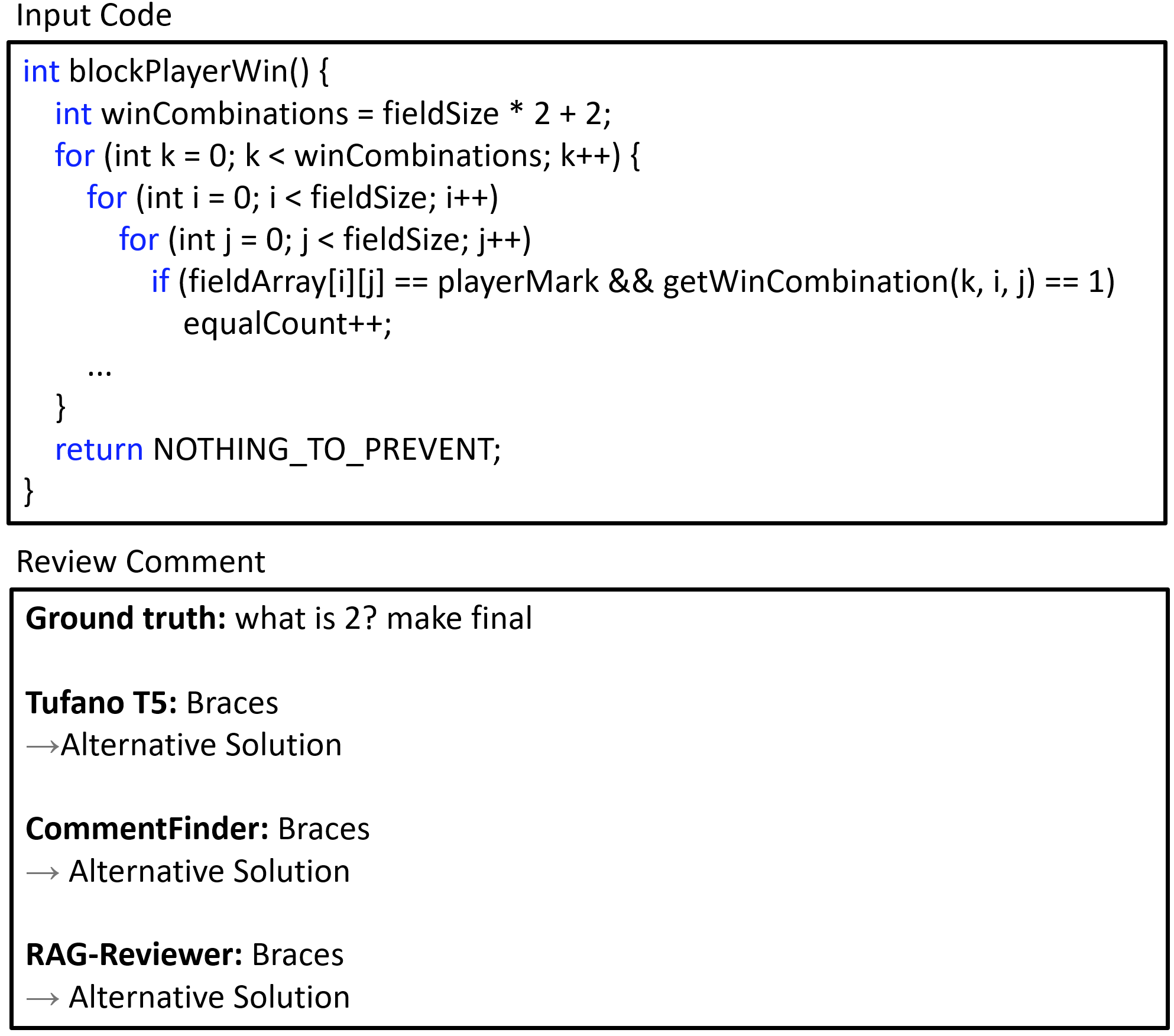}
    \caption{Example of a Alternative Solution review comment from the Tufano et al. manual analysis samples}
    \label{fig:manual_example_alternative}
\end{figure}

Table~\ref{tab:manual_analysis} summarizes the manual classification results. While RAG-Reviewer made fewer exact matches than CommentFinder, it produced more semantically equivalent comments (39 vs. 30) and fewer incorrect ones (50 vs. 57), indicating stronger relevance even when not perfectly matching the ground truth. Compared to Tufano T5, RAG-Reviewer yielded slightly fewer alternative solutions (9 vs. 10) but more semantically equivalent comments (39 vs. 36) and fewer incorrect outputs (50 vs. 54). Figures~\ref{fig:manual_example_semantic} and~\ref{fig:manual_example_alternative} show concrete examples. In Figure~\ref{fig:manual_example_semantic}, RAG-Reviewer generates a semantically correct suggestion regarding \texttt{try-with-resources}, unlike the unrelated outputs from Tufano T5 and CommentFinder. Notably, RAG-Reviewer successfully generates the key LFGT \texttt{try-with-resources} (279 occurrences in the training corpus), which neither Tufano T5 nor CommentFinder produced. Although the input code does not contain this token, RAG-Reviewer can infer it from retrieved exemplars containing relevant context. As shown in Table~\ref{tab:retrieved_exemplars}, several retrieved examples (e.g., top-10, 16, and 20) include \texttt{try-with-resources}, allowing the model to refer to low-frequency but semantically important tokens. This demonstrates how retrieval augmentation provides contextual cues that help the model generalize beyond the input.

\subsection{Future Work}

A key direction for future work is jointly training the retriever and generator. While this study fine-tuned only the generator, prior work\cite{lu2024improving} showed that training both components can improve retrieval quality and overall generation. Another direction is integrating RAG with large language models (LLMs). Although recent studies \cite{nashaat2024towards, lu2023llama} focused on efficient fine-tuning LLMs for RCG, combining them with RAG may further enhance their performance.

\section{Related Work}

In this section, we review related work on Code Review Automation (CRA) from two main perspectives: generation-based and information retrieval-based (IR-based) approaches. We also discuss prior applications of retrieval-augmented generation (RAG) in software engineering.

\subsection{Generation-based Code Review Automation}

With the advancement of pretrained language models (PLMs) \cite{sutskever2014sequence, bahdanau2014neural , vaswani2017attention,devlin2019bert, brown2020language}, there has been growing interest in applying them to various software engineering tasks \cite{feng2020codebert, guo2020graphcodebert, wang2021codet5, guo2022unixcoder}, including CRA. Tufano et al. \cite{tufano2021towards, tufano2022using} were among the first to demonstrate the use of PLMs for CRA by curating a dataset of code–review comment pairs and framing the task as a sequence-to-sequence translation problem. Building on this, Li et al. \cite{li2022codereviewer} proposed specialized pretraining tasks designed to better capture the relationship between code changes (diffs) and corresponding review comments, further improving performance on CRA.

Recent works have continued to explore the use of PLMs on standard CRA benchmarks. For example, Zhou et al. \cite{zhou2023generation} conducted a comparative study and found that CodeT5 \cite{wang2021codet5} consistently outperformed other code-specific PLMs. Beyond PLMs, LLMs have also been adapted to CRA. Lu et al. \cite{lu2023llama} fine-tuned LLaMA \cite{touvron2023llama} using Parameter-Efficient Fine-Tuning (PEFT) techniques and demonstrated competitive performance on the benchmark from Li et al. \cite{li2022codereviewer}, though they reported that Tufano T5 \cite{tufano2022using} still outperformed LLaMA on the Tufano benchmark. Based on this finding, we opted to use PLMs rather than LLMs in our study, as they align better with the Tufano benchmark.

Expanding on Lu et al., Nashaat et al. \cite{nashaat2024towards} introduced a framework that incorporates organizational data with few-shot learning and reinforcement learning from human feedback. More recently, Li et al. \cite{li2025codedoctor} proposed CodeDoctor, a category-aware review comment generator that utilizes LLM-guided classification and a category-specific decoder. While CodeDoctor showed strong performance in multi-category review comment generation, it assumes predefined categories are available and constrains output accordingly. In contrast, our work focuses on general (non-categorical) review comment generation. This direction remains important for building fully automated and flexible CRA systems, as real-world code reviews often include issues that span or fall outside fixed categories.

\subsection{Information Retrieval-based Code Review Automation}

While generation-based CRA shows strong performance, it often incurs high computational cost during training and inference. IR-based CRA offers a more efficient alternative by bypassing language generation and instead reusing comments from similar examples in a retrieval corpus.

Hong et al. \cite{hong2022commentfinder} proposed CommentFinder, an IR-based CRA system that uses sparse Bag-of-Words (BoW) vector representations to retrieve top-k lexically similar code snippets based on cosine similarity. The paired review comments from retrieved code snippets are returned as the predicted output. CommentFinder achieved a 32\% improvement in exact match accuracy and was 49× faster than a generation-based baseline \cite{tufano2022using}. Building on this, Shuvo et al. \cite{shuvo2023recommending} incorporated structural features such as class and library names to improve retrieval quality. More recently, Kartal et al. \cite{kartal2024automating} demonstrated that dense representations from transformer-based models (e.g., UniXCoder \cite{guo2022unixcoder}) outperform sparse representation (e.g., BoW) in CRA retrieval tasks.

\subsection{Retrieval-Augmented Generation in Software Engineering}

RAG combines the strengths of retrieval- and generation-based methods by incorporating retrieved exemplars into the input of a generative model. In software engineering, RAG has shown promise across several tasks, such as code summarization \cite{zhang2020retrieval, parvez2021retrieval, lu2024improving, wei2020retrieve, li2021editsum}, code completion \cite{lu2022reacc}, and automated program repair \cite{jin2023inferfix}.

Zhang et al. \cite{zhang2020retrieval} were among the first to apply RAG to code summarization, using LSTM-based models. Parvez et al. \cite{parvez2021retrieval} extended this idea by using transformer-based PLMs as both retriever and generator. Lu et al. \cite{lu2022reacc} applied RAG to code completion by retrieving similar completed code examples to guide the model. Jin et al. \cite{jin2023inferfix} introduced RAG into LLM-based program repair by augmenting prompts with retrieved bug–fix pairs to improve repair quality.

\section{Threats to Validity}

\textbf{Internal validity.} We reimplemented Tufano T5 in PyTorch due to compatibility issues with its original TensorFlow 2.6.0 code. Pretrained weights were converted accordingly, and the training setup followed prior RCG work \cite{li2022codereviewer, lu2023llama}. While care was taken to align with the original setup, minor discrepancies may exist. Hyperparameters were selected based on prior work and GPU memory limits but may not be optimal.

\textbf{External validity.} Our evaluation is based solely on the Java dataset from Tufano et al. \cite{tufano2022using}. To improve generalizability, future work should evaluate across more languages and diverse software projects.

\section{Conclusion}
In this paper, we presented \textbf{RAG-Reviewer}, a \textbf{RAG}-based code \textbf{Review} comment gen\textbf{er}ation framework. RAG-Reviewer conditions pretrained language models on both input code and retrieved code–review exemplars to improve generation quality. We evaluated the approach on the Tufano et al. benchmark and observed consistent gains over generation-based and IR-based baselines, particularly in generating low-frequency tokens and handling varying input lengths. Our analysis further reveals that using fewer but more informative exemplars—including relevant code—outperforms comment-only augmentation, and that increasing the number of exemplars within the same retrieval strategy leads to additional improvements. These results demonstrate the effectiveness of retrieval augmentation in enhancing code review automation.

\bibliographystyle{IEEEtran}
\bibliography{references}

\end{document}